\documentclass[]{article}

\usepackage{arxiv}

\usepackage[utf8]{inputenc} % allow utf-8 input
\usepackage[T1]{fontenc}    % use 8-bit T1 fonts
\usepackage{hyperref}       % hyperlinks
\usepackage{url}            % simple URL typesetting
\usepackage{booktabs}       % professional-quality tables
\usepackage{amsfonts}       % blackboard math symbols
\usepackage{nicefrac}       % compact symbols for 1/2, etc.
\usepackage{microtype}      % microtypography
\usepackage{lipsum}
\usepackage{graphicx}
\usepackage{tikz}
\usepackage{tabularx} % in the preamble
\usepackage{subcaption}
\usepackage{svg}            % include high-quality vector figures
\usepackage{layouts}        % display layout sizes
\usepackage{comment}        % Use multi-line comments
\usepackage{siunitx}

\title{A nation-wide experiment: fuel tax cuts and almost free public transport for three months in Germany - Report 3 Second wave results}

\author{
Allister Loder \\
Technical University of Munich\\
TUM School of Engineering and Design\\
Chair of Traffic Engineering and Control\\
Arcisstrasse 21, 80333 Munich \\
\texttt{allister.loder@tum.de}\\
\And
Fabienne Cantner\\
Technical University of Munich\\
TUM School of Management\\
TUMCS for Biotechnology \& Sustainability \\ 
Am Essigberg 3, 94315 Straubing\\
\texttt{fabienne.cantner@tum.de}\\
\And
Andrea Cadavid Isaza\\
Technical University of Munich\\
TUM School of Engineering and Design\\
Chair of Renewable and Sustainable Energy Systems\\
Lichtenbergstraße 4a, 85748 Garching\\
\texttt{andrea.cadavid@tum.de}\\
\And
Markus B. Siewert\\
Munich School of Politics and Public Policy\\
TUM Think Tank\\
Richard-Wagner-Straße 1, 80333 München\\
\texttt{markus.siewert@hfp.tum.de}
\And
Stefan Wurster\\
Munich School of Politics and Public Policy\\
Technical University of Munich\\
TUM School of Social Sciences and Technology\\
Professorship of Policy Analysis\\
Richard-Wagner-Straße 1, 80333 München\\
\texttt{stefan.wurster@hfp.tum.de}
\And
Sebastian Goerg\\
Technical University of Munich\\
TUM School of Management\\
TUMCS for Biotechnology \& Sustainability \\ 
Am Essigberg 3, 94315 Straubing\\
\texttt{sebastian.goerg@tum.de}\\
\And
Klaus Bogenberger\\
Technical University of Munich\\
TUM School of Engineering and Design\\
Chair of Traffic Engineering and Control\\
Arcisstrasse 21, 80333 Munich\\
\texttt{klaus.bogenberger@tum.de}\\
}

\begin{document}
\maketitle
\begin{abstract}
In spring 2022, the German federal government agreed on a set of measures that aimed at reducing households' financial burden resulting from a recent price increase, especially in energy and mobility. These measures included among others, a nation-wide public transport ticket for 9\ EUR per month and a fuel tax cut that reduced fuel prices by more than 15\,\%. In transportation research this is an almost unprecedented behavioral experiment. It allows to study not only behavioral responses in mode choice and induced demand but also to assess the effectiveness of transport policy instruments. We observe this natural experiment with a three-wave survey and an app-based travel diary on a sample of hundreds of participants as well as an analysis of traffic counts. In this third report, we provide first findings from the second survey, conducted during the experiment.
\end{abstract}

% keywords can be removed

\section{Introduction}

% uncomment the following line to derive layout sizes for perfectly matching figures
%textwidth in in: \printinunitsof{in}\prntlen{\textwidth}

In transportation research, it is quite unlikely to observe or even perform real-world experiments in terms of travel behavior or traffic flow. There are few notable exceptions: subway strikes suddenly make one important alternative mode not available anymore  \cite{Anderson2014,Adler2016}, a global pandemic changes travelers' preferences for traveling at all or traveling collectively with others \cite{Molloy2021}, or a bridge collapse forces travelers to alter their daily activities \cite{Zhu2010}. However, in 2022 the German federal government announced in response to a sharp increase in energy and consumer prices a set of measures that partially offset the cost increases for households. Among these are a public transport ticket at 9\ EUR per month\footnote{\url{https://www.bundesregierung.de/breg-de/aktuelles/9-euro-ticket-2028756}} for traveling all across Germany in public transport, except for long-distance train services (e.g., ICE, TGV, night trains), as well as a tax cut on gasoline and diesel, resulting in a cost reduction of about 15\ \% for car drivers\footnote{\url{https://www.bundesfinanzministerium.de/Content/DE/Standardartikel/Themen/Schlaglichter/Entlastungen/schnelle-spuerbare-entlastungen.html}}. Both measures were limited to three months, namely June, July and August 2022. As of end of August, more than 52\ million tickets have been sold\footnote{
\url{https://www.vdv.de/bilanz-9-euro-ticket.aspx}
}, while it seems that the fuel tax cut did not reach consumers due to generally increased fuel prices and oil companies are accused of not forwarding the tax cuts to consumers \footnote{\url{https://www.spiegel.de/wirtschaft/tankrabatt-hat-zunehmend-an-wirkung-verloren-rwi-studie-a-cb7a4e84-c943-44a3-b0d3-fcfff9ba3061?dicbo=v2-bf58f4c0d939c05bc696544c175d1063}}. 

For the Munich metropolitan region in Germany, we designed a study under the label "Mobilität.Leben" \footnote{\url{https://www.hfp.tum.de/hfp/tum-think-tank/mobilitaet-leben/}} comprising three elements: (i) a three-wave survey before, during and after the introduction of cost-saving measures; (ii) a smartphone app based measurement of travel behavior and activities during the same period; (iii) an analysis of aggregated traffic counts and mobility indicators. We will use data from 2017 as well as 2019 (pre-COVID-19) and data from shortly before the cost reduction measures for the comparison. In addition, the three-wave survey is presented to a nation-wide control group, which however does not participate in the app. The main goal of the study is to investigate the effectiveness of the cost-saving measures with focus on the behavioral impact of the 9\ EUR-ticket on mode choice \cite{ben1985discrete}, rebound effects \cite{Greening2000,Hymel2010}, and induced demand \cite{Weis2009}. Further details on the study design can be found in our first and second report \cite{reportone,cantner_nation-wide_2022}.

In this third report, we first provide an update on the study participation in Section \ref{sec:participation}; second, we present how participants see and use the 9~EUR-ticket in Section \ref{sec:ticket}; third, Section \ref{sec:mobility} provides a summary of the travel behavior reported in the second wave during the experiment in comparison to the time before; in Section \ref{sec:wtp} we show preliminary results regarding the willingness-to-pay for a successor of the 9~EUR-ticket and finally present the reported impacts on the households' finances in Section \ref{sec:finance}. 

\section{Study participation update} \label{sec:participation}

For the entire study, 2~268 participants had been successfully recruited. The entire sample comprises 1~349 participants for the \textit{Munich study} (MS) and 919 participants from the nation-wide \textit{control group} (CG). At the end of the experiment in August 2022, four participants from the Munich study explicitly opted out. If not stated otherwise, numbers in parentheses refer to the findings from the control group.

The first survey wave has been fully completed by 1~225 participants or 91.1~\% of the recruited participants (CG: 100~\%). The first wave has been distributed to participants at the end of May, right before the beginning of the cost reduction measures in June. The second survey has been fully completed by 1~010 participants or 75.1~\% of the recruited participants (CG: 75.2~\%). The second wave has been distributed at the end of July. In the Munich study 18 participants completed the second survey, but not the first survey, resulting only in 992 total joint responses for the first and second survey or 73.8~\% of the recruited participants. Considering the app participation, at the end of August 2022, 775 participants are still using the app of which 717 have completed both surveys.

\section{9~EUR-ticket} \label{sec:ticket}

In the Munich study, 90~\% (CG: 49~\%) and 89~\% (CG: 48~\%) of the second-wave respondents bought the 9~EUR-ticket for June and July, respectively. For August, 49~\% (CG: 20~\%) bought the 9~EUR-ticket and 40~\% (CG: 29~\%) had the intention to buy it when asked at the end of July. Consequently, we can conclude that the interest in holding this travel pass did not change over time, while the data suggests that interest on the other hand did not increase over time. Officially, around 52 million 9~EUR-tickets have been sold in total during the three months and ten million people received it indirectly through their existing travel card subscriptions. This corresponds to around one third of Germany's population. Consequently, the ownership shares in the Munich study and the control group are remarkably higher.

In the second wave, the 9~EUR-ticket receives support from respondents across the board. For example, 84~\% (CG: 61~\%) of respondents agree to the statement that the 9~EUR-ticket leads to a more comprehensible pricing structure, while 85~\% (CG: 72~\%) agree to the statement that the new travel makes traveling in Germany more flexible. 80~\% of respondents (CG: 60~\%) agree that the 9~EUR-ticket makes them less worried to buy the wrong public transport ticket. Regarding the savings caused by the 9~EUR-ticket, 64~\% (CG: 65~\%) of respondents agree to the statement that the savings can be spent for meaningful products or services.

\section{Travel behavior} \label{sec:mobility}

\begin{figure}
    \centering
    \includegraphics[width=14cm]{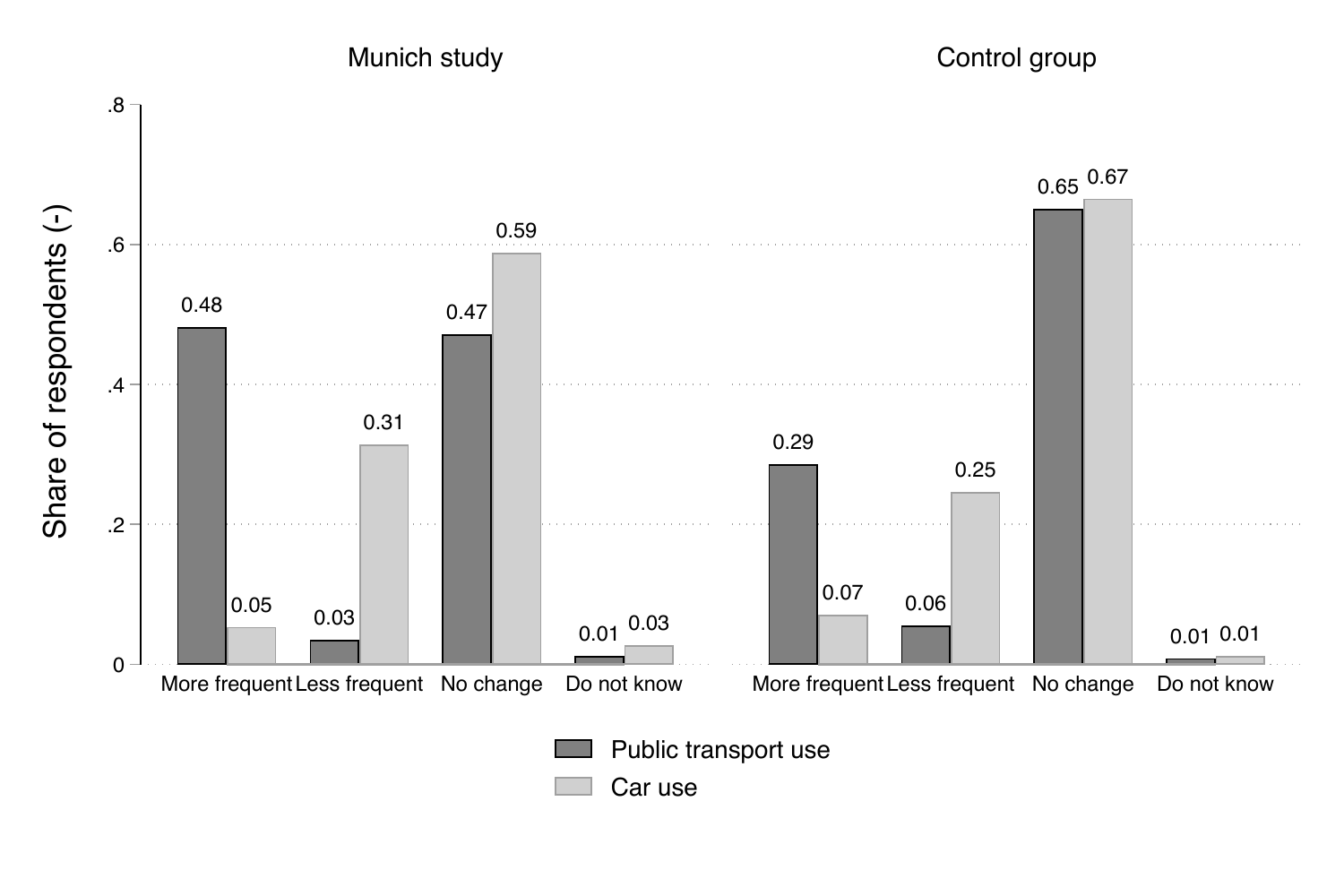}
    \caption{Stated changes in travel behavior during the period of the 9~EUR-ticket and the fuel tax cut compared to the time before.}
    \label{fig:travel}
\end{figure}

When asked about their behavioral changes during the first weeks of the 9~EUR-ticket and fuel rebate compared to the time before, 48~\% (CG: 29~\%) of respondents state as seen in Figure \ref{fig:travel} that they increased public transport use, 3~\% (CG: 6~\%) state that they decreased public transport use, and 47~\% (CG: 65~\%) report no change. Regarding car use, 5~\% (CG: 7~\%) of respondents state that they increased car use, 31~\% (CG: 25~\%) that they decreased car use and 59~\% (CG: 67~\%) state no change in their car use. Overall, 89~\% (CG: 87~\%) of respondents who increased public transport use state that this was in response to the introduction of the 9~EUR-ticket, while only 74~\% (CG: 66~\%) of respondents who decreased car use state that this was in response to the introduction of the 9~EUR-ticket. Interestingly, regarding the increase in car use, 2~\% (CG: 40~\%) of respondents argue that their increase is in response to the fuel tax cut. 

Considering both modes jointly, we find that 26~\% (CG: 17~\%) of respondents report an increase in public transport use and an decrease in car use. Out of these, more than 82~\% (CG: 84~\%) of respondents argue that the reason for their behavioral change is the introduction of the 9~EUR-ticket.

We corroborate these first findings by considering the stated weekly car and public transport usage patterns reported in the first and second wave. Here, we classify usage as follows: never, less than once per week, once per week, 2-3 days per week, 4-5 days per week and 6-7 days per week. Thus, observing a change here, is likely to be more robust. We find that 55~\% (CG: 63~\%) of respondents did not change their stated car use, while 46~\% (CG: 58~\%) of respondents did not change their stated public transport use; 24~\% (CG: 18~\%) of respondents show an increase and 22~\% (CG: 18~\%) of respondents a decrease in car use accordingly; 37~\% (CG: 23~\%) of respondents show an increase and 17~\% (CG: 18~\%) show a decrease in public transport use respectively. 

Considering both modes jointly, we find that 8.6~\% (CG: 5.9~\%) of respondents increased public transport and decreased car use. When focusing only on those respondents who used the car at least four days per week, i.e. those who can change their behavior, we find that 18~\% (CG: 7.2~\%) of respondents increased public transport and decreased car use. The substantial difference between this figure and the 26~\% (CG: 17~\%) reported above can be explained by the fact that the lower figure refers to a more coarse weekly pattern classified into bins of two days width. This means that, for example, replacing one out of two car trips per day by public transport is not reflected in this classification, only if it is completely abandoned for one day or more. However, the weekly pattern can be expected to provide more robust estimates regarding the impact of the behavioral change, e.g., in terms of kilometers traveled.  

Regarding activities, 35~\% (CG: 23~\%) of respondents state that they participated in more activities as a consequence of the 9~EUR-ticket, while 44~\% (CG: 59~\%) state that the introduction of the new travel pass did not increase their number of activities. Respondents report to use public transport on average for 1.2 more activities per week and to use the car for 0.7 less activities per week. This finding again provides evidence that the introduction of the 9~EUR-ticket generates to some extent new travel demand.

\section{Willingness-to-pay for a successor of the 9~EUR-ticket} \label{sec:wtp}

\begin{figure}[t!]
    \centering
     \begin{subfigure}[b]{0.49\textwidth}
         \centering
         \includegraphics[width=\textwidth]{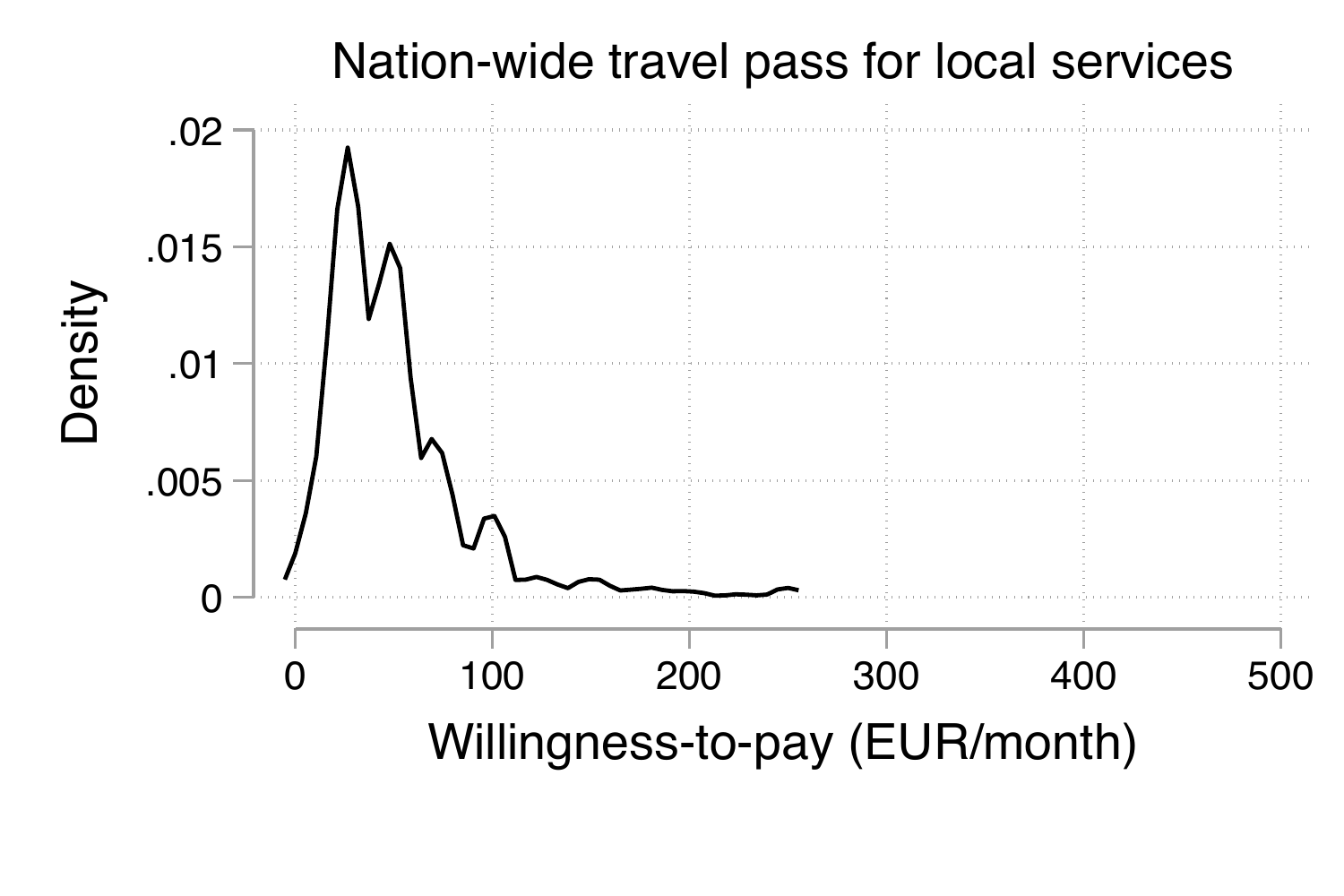}
         \caption{}
         \label{fig:local}
     \end{subfigure}
     \hfill
     \begin{subfigure}[b]{0.49\textwidth}
         \centering
         \includegraphics[width=\textwidth]{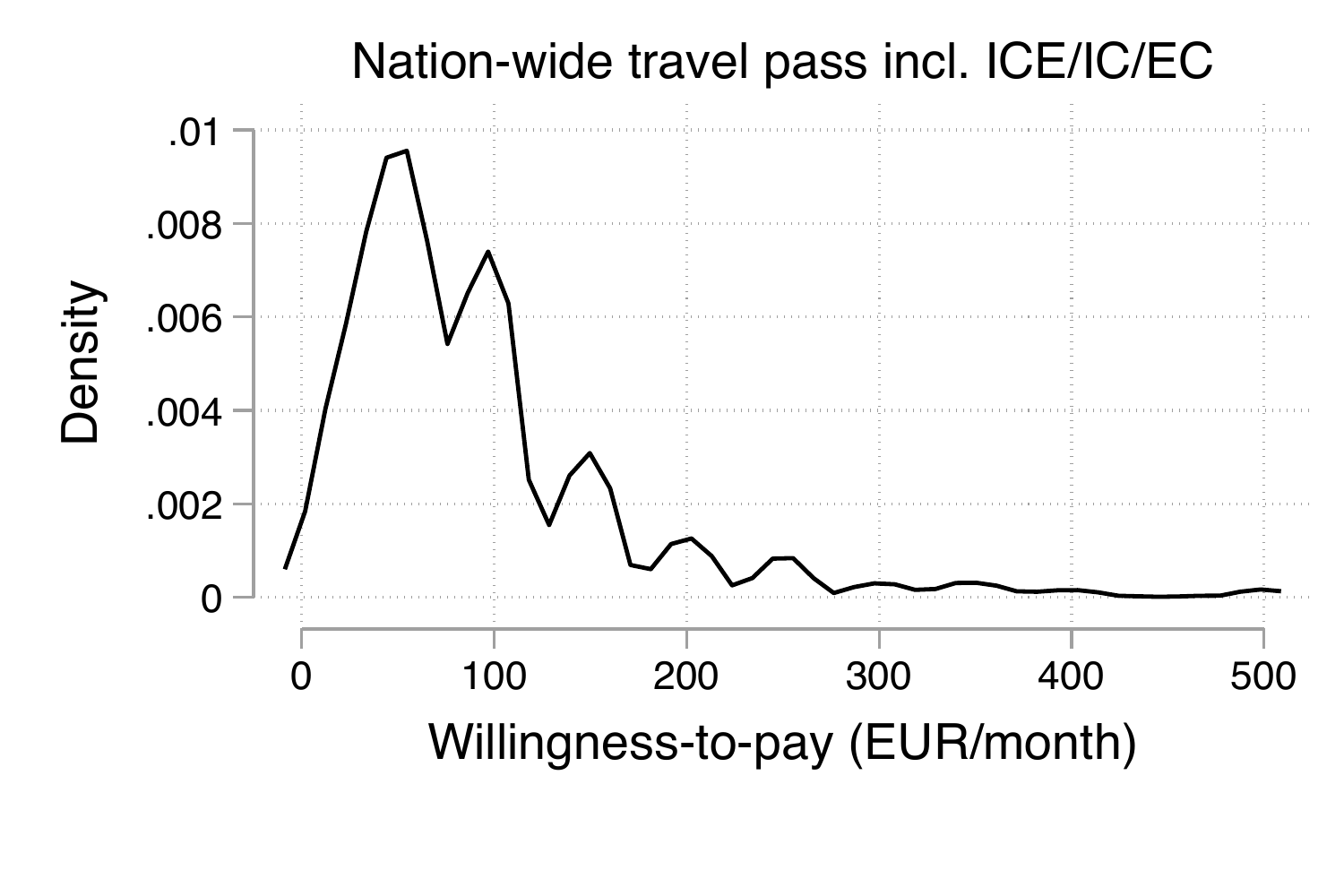}
         \caption{}
         \label{fig:nation}
     \end{subfigure}
    \caption{Willingness-to-pay for nation-wide travel passes. (a) shows the distribution for a travel pass for local public transport services only and (b) shows the distribution for a travel pass for all public transport services, incl. long-distance services like ICE, IC and EC.}
    \label{fig:wtp}
\end{figure}

In the second survey, respondents were asked to state their maximum willingness-to-pay for a nation-wide travel pass for all local public transport services and for all public transport services including long-distance services. Figure \ref{fig:wtp} shows the distributions of the responses. The average willingness-to-pay for a nation-wide travel pass for all local public transport services as a successor of the 9~EUR-ticket was 52.39~EUR (CG: 47.74~EUR), while the average willingness-to-pay for a nation-wide travel pass for all public transport services including long-distance services was 101.47~EUR (CG: 77.62~EUR). 

For the nation-wide travel pass for all local services, we find that higher incomes increase the willingness-to-pay by 10~EUR to 15~EUR compared to the lowest income group. We find no statistically significant differences between males and females, but a small age-effect of about minus 2~EUR per ten years of age in the willingness-to-pay, i.e., older people are willing to spend less for such a ticket. Students willingness-to-pay is about 7~EUR lower compared to working people. Using the car frequently does not impact the willingness-to-pay, but being a public transport user before the introduction of the 9~EUR ticket increases the willingness-to-pay by around 18~EUR. 

The tendency of effects is similar for the nation-wide travel pass for all services including long-distance services, but at a higher level as it can be seen in Figure \ref{fig:wtp}. However, for this type of travel pass an effect of gender exist as women's willingness-to-pay is around 8~EUR less. 

Comparing the findings from Figure \ref{fig:wtp} to the public debate we find that our sample's average willingness-to-pay is very close to the discussed 49~EUR, which has been proposed by the Social Democratic and the Green Party, currently part of the federal government. Contrary, the willingness-to-pay is substantially below the 69~EUR as proposed by the Association of German Transport Companies (VDV). 

\section{Financial aspects} \label{sec:finance}

The primary intention of the fuel tax cut and the 9~EUR-ticket has been to partially offset the recent price increases. Therefore, it is not surprising that the respondents' agreement to the statement that the ticket is an relief stays high at 82~\% (CG: 75~\%). Overall, 76~\% (CG: 39~\%) of respondents state that they benefited financially from the 9~EUR-ticket, while only 24~\% (CG: 32~\%) mentioned that they benefited financially from the fuel tax cut. 

Table \ref{tab:1} shows the stated savings of the Munich study respondents (and the control group in parentheses). It can be seen that the savings for most respondents are less than 50~EUR per month. Yet, a substantial portion stated savings between 50~EUR and 100~EUR per month. Overall, 48.9~\% (CG: 55.2~\%) of respondents state that these savings have already been consumed by inflation, while 14.5~\% (CG: 11.7~\%) state that they used these savings for spending on other goods and services.

\begin{table}[h!]
\centering
    \begin{tabularx}{\textwidth}{l | cccccc} 
    \toprule
    & \multicolumn{6}{c}{Savings per month in EUR}\\ \cmidrule{2-7}
     & < 50  &  50-100  &   100-150  &  150-200  &   200-250  &  > 250  \\
    \midrule
    Benefit from ticket, N = 781 (269) & 45.1 (40.9) & 38.7 (36.4) & 10.8 (13.4) & 4.1 (6.7) & 0.8 (1.9) & 0.6 (0.7) \\
    Benefit from fuel tax, N = 238 (221) & 47.0 (56.11) & 35.3 (26.7) & 11.3 (11.3) & 5.0 (4.5) & 0.4 (1.4) & 0.8 (0.0) \\
    Benefit from both, N = 193 (101) & 41.5 (34.7) & 38.9 (37.6) & 13.0 (16.8) & 5.2 (8.9) & 0.5 (2.0) & 1.0 (0.0) \\
    \toprule
    \end{tabularx}
    \caption{Distribution of respondents across savings per month from the fuel tax cut and the 9~EUR-ticket in percent. The values from the control group are given in parentheses.}
    \label{tab:1}
\end{table}

% -----------
\section{Discussion and outlook}

In this third report, we have provided some first insights into the second wave of our study "Mobilität.Leben". Based on the responses it can be seen that the introduction of the 9~EUR-ticket impacted mobility and everyday life. There is evidence that some respondents changed from the car to public transport, but in order to estimate the precise extent of this effect we need to more data and analyses from our app-based travel diary. There is also some evidence that the 9~EUR-ticket increased travel demand. Nevertheless, the first findings indicate similar effect sizes as already found elsewhere \cite{keblowski_why_2020}.

In closing, it should be noted that this report does not present the final results of our study and are therefore preliminary. The presented results are not yet weighted to correspond to a representative sample. Thus, the findings at this moment in time only describe our sample from the Munich metropolitan area and the control group. The next steps include the analysis of travel behavior based on the smartphone app before, during and after the fuel tax cut and the 9~EUR-ticket as well as the completing the study with the third survey wave in September and October, which includes a stated preference experiment on the pricing of a successor ticket of the 9~EUR-ticket.

\section*{Acknowledgements}

The authors would like to thank the TUM Think Tank at the Munich School of Politics and Public Policy led by Urs Gasser for their financial and organizational support and the TUM Board of Management for supporting personally the genesis of the project. The authors thank the company MOTIONTAG for their efforts in producing the app at unprecedented speed. Further, the authors would like thank everyone who supported us in recruiting participants, especially Oliver May-Beckmann (M Cube) and Ulrich Meyer (TUM), respectively.

\bibliographystyle{unsrt}  
\bibliography{references}  %%% Remove comment to use the external .bib file (using bibtex).

%%% Comment out this section when \bibliography{references} is enabled.
%\begin{thebibliography}{1}

%\end{thebibliography}

\end{document}